\documentclass{ws-procs975x65}
\usepackage{graphicx}

\def\beq{\begin{equation}}
\def\eeq{\end{equation}}
\begin{document}

\title{Evidence for Planck-scale resonant particle production during 
inflation from the CMB power spectrum}
\author{Grant J. Mathews$^*$ and Mayukh R. Gangopadhyay}

\address{Department of Physics, University of Notre Dame,\\
Notre Dame, IN 46556, USA\\
$^*$E-mail: gmathews@nd.edu}

\author{Kiyotomo Ichiki} 

\address{Department of Physics, Nagoya University, Nagoya 464-8602, Japan}

\author{Toshitaka Kajino}

\address{National Astronomical Observatory of Japan\\
Mitaka, Tokyo 181-8599, Japan}

\begin{abstract}
 The power spectrum of the cosmic microwave background from both the {\it Planck} and {\it WMAP} data exhibits a slight dip  for multipoles in the range of
 $l= 10-30$.  We show that such a dip could be the result of the resonant creation of  massive particles that couple
 to the inflaton field.   For our
best-fit models, the epoch of resonant particle creation reenters the
horizon at a wave number of $k_* \sim 0.0011 \pm 0.0004 $  ($h$
Mpc$^{-1}$).  The amplitude and location of this feature corresponds
to the creation of a number of degenerate fermion species of mass $\sim (8-11) /\lambda^{3/2} $ $m_{pl}$ during
inflation where $\lambda \sim (1.0 \pm 0.5) N^{-2/5}$ is the coupling constant between the inflaton field and the
created fermion species, while $N$ is the number of degenerate species.  Although the evidence is of
marginal statistical significance,  this could constitute  new observational hints of unexplored physics beyond the Planck scale.\end{abstract}

\keywords{Inflation; cosmic microwave background; string theory}

\bodymatter

%%%%%%%%%%%%%%%%% now a standard article style for the most part

\section{Introduction}

 In this paper we summarize an analysis\cite{Mathews15} of a peculiar feature visible in the observed power spectrum 
near multipoles $\ell = 10-30$.  This is an interesting region in the CMB power spectrum because it corresponds to 
angular scales that are not yet in causal contact, so that the observed power spectrum is close to  the true primordial power spectrum.  

\begin{figure}[htb]
\includegraphics[width=3.0in,clip]{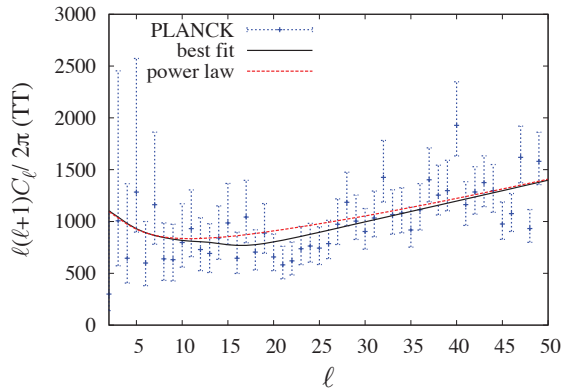} 
\caption{(Color online) CMB power spectrum in the range of $\ell = 3-50$  Points with error bars are from the 
{\it Planck} Data Release \cite{PlanckXIII}.  The dashed line shows the best standard $\Lambda$CDM fit to the {\it Planck} CMB power spectrum
based upon a power-law primordial power spectrum.  The solid line shows the best fit for a model with resonant particle creation during inflation.}
\label{fig:1}
\end{figure}

The {\it Planck}\cite{PlanckXIII} observed power spectrum in this region is shown in Figure \ref{fig:1} from Ref.~[\refcite{Mathews15}].
Although the error bars are large, there is a noticeable systematic deviation  in the range $\ell = 10-30$ below the  best fit based upon the standard $\Lambda$CDM cosmology with a power-law primordial power spectrum.  There is also a well-known suppression of the quadrupole moment in the CMB.
These same  features are  visible in the CMB power spectrum from the Wilkinson Microwave Anisotropy Probe ({\it WMAP}) \cite{WMAP9}, and hence, are likely a true feature in the CMB power spectrum, although it should be noted that in the Planck 
Cosmological parameters paper,\cite{PlanckXX}  
the deviation from a simple power law  in the range $\ell = 10-30$ was deduced to be of  weak  statistical significance due to the large cosmic variance at low $\ell$.  

A  number of mechanisms have been proposed (see summary in Ref.~[\refcite{Mathews15}]) to deal with the suppression of the power spectrum on large scales and low multipoles.   Most of these works, however, were concerned with the suppression of the lowest moments via changes in the inflation generating potential.  
  In the present work\cite{Mathews15}, however, we are concerned specifically with suppression of the power spectrum in the range $\ell = 10-30$ due
  to the  possibility that  new trans-Planckian physics occurs  near the end of the
inflation epoch corresponding to  the resonant creation \cite{chung00,Mathews04} of Planck-scale particles that
 couple to the inflaton field.  Our best fit is  shown  by the solid line in Figure \ref{fig:1} which  we describe in detail in the following sections.

This interpretation has the
intriguing aspect that, if correct, an opportunity emerges to use the
CMB  to probe  properties of new particle species that existed  at and above the  Planck scale ($m_{pl} \sim
10^{19}$ GeV).  That is the goal of the present work.
Indeed, string theory compactification schemes generically
postulate the existence of massive particles at or above the Planck
scale from the Kaluza-Klein states,
winding modes,  string excitations, etc.
Moreover, the coupling of the inflaton to other particle species near the end of inflation is not only natural, but probably required.  
This is because the energy density in the inflaton must be converted to entropy in light or heavy particle species at the end of inflation as a means to reheat the universe.
 Hence, the existence of Planck-scale
mass particles that couple to the inflaton near the end of inflation is a scenario that is both natural and even required.  Moreover, this provides a possible opportunity to uncover new physics in the trans-Planckian regime.

\section{Resonant Particle Production during Inflation}

The details of the  resonant particle creation paradigm during inflation have been explained in Refs.~\refcite{Mathews15,chung00,Mathews04}.  Indeed, the idea was originally introduced \cite{Kofman94} as a means for reheating after inflation.
Since \cite{chung00} subsequent work \cite{Elgaroy03, Romano08, Barnaby09, Fedderke15}  has elaborated on the basic scheme into a model with coupling between two scalar fields.  

In this minimal extension from the basic picture, the
inflaton $\phi$ is postulated to couple to particles
whose mass is of order the inflaton field value.  These particles
are then resonantly produced as the field obtains a critical value
during inflation.  If even a small fraction of the 
inflaton field is affected  in this way, it can produce an observable feature in
the primordial power spectrum. In particular, there can  be either an excess in the power spectrum as noted in \cite{chung00,Mathews04},
or a dip in the power spectrum as described in this paper. Such a dip offers important new clues to the trans-Planckian physics of the early universe.

In the simplest slow roll approximation \cite{Liddle,cmbinflate}, the
generation of density perturbations of  amplitude,
$\delta_H(k)$, when  crossing the Hubble
radius is just,
\begin{equation}
\delta_H(k) \approx {H^2
\over 5 \pi \dot \phi}~~,
\label{pert}
\end{equation}
where $H$ is the expansion rate, and $\dot \phi$ is the rate of change of
the inflaton field when the comoving wave number $k$ crosses the
Hubble radius during inflation.  

For the application here,\cite{Mathews15} we adopt a  positive Yukawa coupling of strength $\lambda$ between  an inflaton field $\phi$ and a field $\psi$ of
$N$  degenerate fermion species.The total Lagrangian density including the inflaton scalar field $\phi$, the Dirac fermion field, and
the Yukawa coupling term is simply,
\begin{eqnarray}
{\cal L}_{\rm tot} &=& \frac{1}{2}\partial_\mu \phi  \partial^\mu \phi - V(\phi) \nonumber \\
&+& i \bar \psi  \gamma^\mu \partial_\mu \psi - m \bar \psi  \psi + N \lambda \phi \bar \psi  \psi ~~.
\end{eqnarray}
For this Lagrangian, it is obvious that the fermions have an effective mass of
\begin{equation} 
M(\phi) = m - N \lambda \phi~~.
\end{equation}
This  vanishes
for a critical value of the inflaton field,
$\phi_* = m/N \lambda$.  Resonant fermion production
will then occur\cite{chung00} in a narrow range of the inflaton field amplitude
around $\phi = \phi_*$.

The cosmic scale factor is labeled $a_*$ at the
time $t_*$ at which resonant particle production occurs.  Considering
a small interval around this epoch, one can treat $H = H_*$ as
approximately constant (slow roll inflation).  The number density $n$
of particles can be taken as zero before $t_*$ and afterwards as $n =
n_*[a_*/a(t)]^{3}$.  The fermion vacuum expectation value can then be
written,
\begin{equation}
 \langle \bar \psi \psi \rangle = n_* \Theta (t-t_*) \exp{[-3 H_*(t-t_*)]} ~~.
 \end{equation}
where $\Theta$ is a step function.

Then,  one obtains\cite{Mathews15}  the perturbation in the primordial power spectrum as it exits the horizon:
\begin{equation}
\delta_H = \frac{[\delta_H(a)]_{N \lambda = 0}}{1 + \Theta (a - a_*)( N \lambda n_*/\vert \dot \phi_*\vert H_*) (a_*/a)^3 \ln{(a/a_*)}} ~~.
\label{deltahnew}
\end{equation}
It is clear that the power in the fluctuation of the inflaton field will diminish as the particles are resonantly created when the universe
grows to some critical scale factor $a_*$.

Using $k_*/k = a_*/a$, then
the perturbation spectrum Eq.~(\ref{deltahnew})
can be reduced \cite{Mathews15} to a simple 
two-parameter function.
\begin{equation}
\delta_H (k) = \frac{[\delta_H(a)]_{N \lambda = 0}}{1 + \Theta (k-k_*)A (k_*/k)^3 \ln{(k/k_*)}} ~~.
\label{perturb}
\end{equation}
where the amplitude  $A$ and characteristic wave number $k_*$ ($k/k_*\ge 1$) can be fit to the observed power spectrum from the relation:
$k_* = { \ell_* }/{ r_{lss}}$, 
 where $r_{lss} $ is the
comoving distance to the last scattering surface, taken here to be 14 Gpc.
The values of $A$ and $k_*$ determined from from the CMB power spectrum
relate to the inflaton coupling $\lambda$ and fermion mass $m$, for a
given inflation model via Eqs.~(\ref{deltahnew}) and (\ref{perturb}). 
\begin{equation}
A  = |\dot{\phi}_*|^{-1} N \lambda
n_* H_*^{-1} ~~.
\end{equation}

%We have MCMC fit the Planck CMB power spectrum using the perturbation of Eq.~(\ref{perturb}) 
%plus a a standard power-law $k^n$ primordial power spectrum with other cosmological parameters fixed at the maximum likelihood values from the Planck analysis..
 The connection between 
resonant particle creation and the CMB temperature fluctuations 
is straightforward.  We have made a  multi-dimensional 
 Markov Chain Monte-Carlo
analysis \cite{Christensen,Lewis} of the 
CMB using the {\it Planck}  data \cite{PlanckXIII} and the {\it CosmoMC} code \cite{Lewis}.   
For simplicity and speed in the present study we
only marginalized over 
parameters which do not alter the matter or CMB transfer functions. Hence, we only varied $A$ and $k_*$, along with the six parameters, 
 $\Omega_b h^2, \Omega_c h^2, \theta, \tau, n_s, A_s$.  Here,  $\Omega_b $ is the baryon content, $ \Omega_c $ is the cold dark matter content,  $\theta$ is the acoustic peak angular scale, $\tau$ is the optical depth, $n_s$ is the power-law spectral index, and $ A_s$ is the normalization.
%Hence, the set of free parameters in the analysis is ($n_s, A_s, \log{(k_*)},
%A,_* $), where $n_s$ is the spectral index, 
%$A_s$ is the overall amplitude of the primordial power spectrum,
%As usual,
%both $n_s$ and $A_s$ are normalized 
%at $k = 0.05$ Mpc$^{-1}$. 

%Figure \ref{fig:2} shows contours of likelihood for the resonant particle creation parameters,
%$A$ and $k_*$.  
Adding this perturbation to the primordial power spectrum improves the total 
$\chi^2$ for the fit from 9803 to 9798.  One expects that the effect of interest here would only make a small change ($\Delta \chi^2 = 5$) in the overall fit because  
it only affects a limited range of $l$ values with large error bars.  Nevertheless, from
 the likelihood contours we can deduce a mean value of $A = 1.7 \pm 1.5$ with a maximum likelihood value of   $A = 1.5$,  and a mean value of $k_* = 0.0011\pm 0.0004 ~h~{\rm Mpc}^{-1}$

%\begin{figure}[htb]
%\includegraphics[width=3.5in,clip]{dip_2D_.pdf} 
%\caption{(Color online) Constraints on parameters $A$ and $k_*$ from the
%MCMC analysis of the CMB power spectrum.
%Contours show 1 and  2$ \sigma$ limits.  The horizontal
%axis is in units of ($h$ Mpc$^{-1}$). }
%\label{fig:2}
%\end{figure}

%Of course, it is obvious that adding extra parameters should improve the goodness of fit.  One should quantify the statistical significance of the improvement over a simple power-law primordial power spectrum.  A $\Delta \chi^2 = 5$ in the fit corresponds to a 92\% confidence level for two free parameters, hence less than a 2$\sigma$ confidence limit.  To be more precise, the Bayesian information criterion (BIC) can be used to select whether one model is better than another by introducing a penalty term for the number of parameters in the model fit.
%Under the assumption that the model errors are independent and obey a normal distribution, then the BIC can be rewritten in terms of $\Delta \chi^2$ as 
%BIC$ \approx \Delta \chi^2  + df \cdot \ln{n}$ where $df$ is the number of degrees of freedom in the test and $n$ is the number of points in the observed data. For the 30 multipoles in the range of the fit,  the introduction of 2 new free parameters then corresponds to a BIC$ = 1.8$. Generally, BIC$> 2$ is considered positive evidence for an improvement in the fit.  Hence, one must conclude that the evidence for this fit is statistically weak.  Nevertheless, it is worthwhile to examine the possible physical meaning of the deduced parameters.

\section{Physical Parameters}

The coefficient $A$
can be related\cite{Mathews15}  directly to the coupling constant $\lambda$ 
This gives 
\begin{equation}
A \sim 1.3 N  \lambda^{5/2}.
\end{equation}
Hence, for the maximum likelihood value of $A \sim 1.5 $, we have 
\begin{equation}
\lambda  \approx \frac{(1.0 \pm 0.5)}{N^{2/5}} ~~.  
\label{Nconst}
\end{equation}
 So,   $\lambda \le 1$ requires $N>1$ as expected.

The fermion particle mass $m$ can then be deduced from $m = N \lambda \phi_*$.  From Eq.~(\ref{Nconst}) then we have
$m \approx   \phi_*/\lambda^{3/2}$.
For this purpose, we adopt a general monomial potential for which:
\begin{equation} 
\phi_* = \sqrt{2 \alpha {\cal N}} m_{pl}~~.
\end{equation}

For $k_* = 0.0011\pm 0.0004 ~h~{\rm Mpc}^{-1}$, and $k_H = a_0 H_0 = (h/3000)$ Mpc$^{-1} \sim 0.0002,$ we have ${\cal N} - {\cal N}_*  = ln{(k_H/k_* )} <1$.  Typically one expects  ${\cal N}(k_*) \sim  {\cal N} \sim 50 - 60$.  We note, however, that one can have the number of e-folds as low as ${\cal N} \sim 25$ in the case of thermal inflation \cite{Liddle}.  For standard inflation a monomial potential with   $\alpha = 2$ would have $\phi_* \sim (14- 15)~ m_{pl}$.  However, the limits on the tensor to scalar ration from the {\it Planck} analysis \cite{PlanckXX} rule out $\alpha = 2$ at the 95\% confidence level.  Monomial potentials are more consistent with  $\alpha = 1$ ($\phi_* = (10-11) ~m_{pl}$), or even $\alpha = 2/3$ ($\phi_* = (8-9) ~m_{pl}$).  
Hence,  we have roughly the constraint,  
\begin{equation}
m \sim  (8-11)  ~\frac{m_{\rm pl}}{\lambda^{3/2}}~~.
\end{equation}
So, one can deduce\cite{Mathews15} a family of possible  properties of the resonantly produced particle (i.e. its mass and coupling strength)  in terms of a single parameter, the degeneracy $N$.
%This is illustrated in Figure \ref{fig:3} that shows allowed values and uncertainty in the coupling constant and particle mass as a function of the number of degenerate species for a $\phi^{2/3}$ inflaton effective potential experiencing 50 $e$-folds of inflation.     

%\begin{figure}[htb]
%\includegraphics[width=3.5in,clip]{partc-m-lam.pdf}
% \caption{ Implied values of $\lambda$ and $m$ as a function of the number of degenerate species $N$ for a $\phi^{2/3}$ inflaton effective potential experiencing 50 $e$-folds of inflation. Upper curves show the allowed mass.  Lower curves are the coupling constant.  Solid line is for the best-fit normalization $A = 1.5$.  The dashed and dot-dashed lines show the uncertainties due to the upper and lower limits on $A$, respectively.}
%\label{fig:3}
%\end{figure}

\section{Conclusion}
We have analyzed the $\ell = 10-30$ dip in the {\it Planck} CMB power spectrum in the context of a
model for the creation of $N$ nearly degenerate  trans-Planckian massive fermions during inflation.   The  best fit to the  CMB  power spectrum
implies an optimum feature at $k_* = 0.0011 \pm 0.0004~h$\,Mpc$^{-1}$ and  $A \approx  1.7 \pm 1.5$.  For monomial inflation potentials consistent with the {\it Planck} tensor-to-scalar ratio,  this feature  would correspond to the resonant creation
of nearly degenerate  particles with $m \sim 8-11$ $m_{pl}/\lambda^{3/2}$ and a Yukawa coupling
constant $\lambda$ between the fermion species and the inflaton field of $\lambda
\approx (1.0 \pm 0.5)N^{-2/5}$ for $N$  degenerate fermion species.
 
 We conclude that if the present
analysis is correct, this may be one of the first  hints at observational
evidence of new particle physics at the Planck scale.  Indeed, one
expects a plethora of particles at the Planck scale, particularly in
the context of string theory.  Perhaps, the presently observed CMB power
spectrum contains the first suggestion that a subset of such particles may have 
 coupled to the inflaton field leaving a relic signature of their existence in the
CMB primordial power spectrum.

\section*{Acknowledgments}
Work at the University of Notre Dame is supported
by the U.S. Department of Energy under 
Nuclear Theory Grant DE-FG02-95-ER40934.
Work at NAOJ was supported in part by Grants-in-Aid for Scientific Research of JSPS (26105517, 24340060).
Work at Nagoya University supported by JSPS research grant number 24340048.


\begin{thebibliography}{0}
%
\bibitem{Mathews15} G. J. Mathews, M. R.  Gangopadhyay, K. Ichiki, and T. Kajino, Phys. Rev. {\bf D92} 123519 (2015).
%
\bibitem{PlanckXIII} {\it Planck} Collaboration, Astron. \& Astrophys. Submitted (2015) ArXive:1502.01589, astro-ph
%
\bibitem{WMAP9} G. Hinshaw,  et al. ({\it WMAP Collaboration})    Astrophys. J. Suppl. Ser., {\bf 208}, 19 (2013).
%
\bibitem{PlanckXX} {\it Planck} Collaboration, Astron. \& Astrophys. Submitted (2015) ArXive:1502.02114, astro-ph
%
%
\bibitem{chung00} D. J. H. Chung, E. W. Kolb, A. Riotto, and I. I. Tkachev, Phys. Rev. D {\bf 62}, 043508 (2000).
%
\bibitem{Mathews04}G. J. Mathews, D.  Chung, K. Ichiki, T. Kajino, and M. Orito,  Phys. Rev. {\bf D70},  083505 (2004). 
%
%
%\bibitem{Chung12} D. J. H. Chung, L. L. Everett, H. Yoo, P. Zhou, Phys. lett. {\bf B712}, 147.
%
%\bibitem{Battefeld} D. Battefeld, T. Battefeld, and D. Fiene,  Phys. Rev., {\bf D89}, 123523 (2014).
%
\bibitem{Kofman94} L. Kofman, A. D. Linde, and A. A. Starobinsky, Phys. Rev. Lett. {\bf 73} 3195 (1994).
%
\bibitem{Elgaroy03} O. Elgaroy, S. Hannestad, and T. Haugboelle, JCAP, 09, 008 (2003).
%
\bibitem{Romano08} A. E. Romano and M. Sasaki, Phys. Rev. D {\bf 78}, 103522 (2008).
%
\bibitem{Barnaby09} N. Barnaby, Z. Huang, L. Kofman, and D. Pogosyan, Phys. Rev. D {\bf 80}, 043501 (2009).
%
\bibitem{Fedderke15} M. A. Fedderke, E. W. Kolb, M. Wyman, Phys. Rev.,  D {\bf 91}, 063505 (2015).
%
\bibitem{Liddle}
A. R. Liddle and D. H. Lyth, {\it Cosmological Inflation and Large 
Scale Structure}, (Cambridge University Press: Cambridge, UK), (1998).
%
\bibitem{cmbinflate}
E. W. Kolb and M. S. Turner, 
{\it The Early Universe}, (Addison-Wesley, Menlo Park, Ca., 1990).
%
%\bibitem{Starobinsky02} A. A. Starobinsky and I. I. Tkachev, J. Exp. Th. Phys.  Lett., {\bf 76}, 235 (2002).
%
%\bibitem{Camb}
%A. Lewis, A. Challinor,  and A. Lasenby, Astrophys. J., {\bf 538}, 473 (2000).
%
\bibitem{Christensen}
N. Christensen and R. Meyer, L. Knox, and B. Luey,  
Class. and Quant.  Grav., {\bf 18}, 2677 (2001).
%
\bibitem{Lewis}
A. Lewis and S. Bridle, Phys. Rev. D {\bf 66}, 103511 (2002).
%
\bibitem{birrellanddavies} N. D. Birrell and P. C. W. Davies, Quantum
Fields in Curved Space, (Cambridge Univ. Press, Cambridge, 1982).
%
\end{thebibliography}
\end{document}